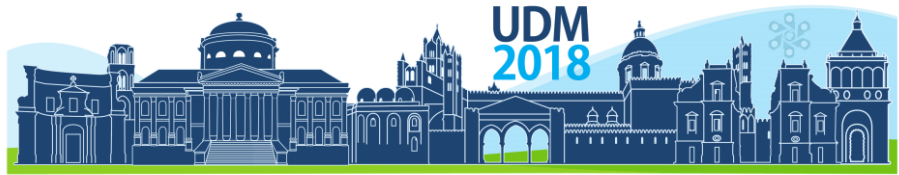

# Modelling overland flow from local inflows in "almost no-time" using Self-Organizing Maps

**João P. Leitão[1,*], Mohamed Zaghloul[2,3] and Vahid Moosavi[2]**
[1] Eawag: Swiss Federal Institute of Aquatic Science and Technology, Department Urban Water Management, Dübendorf, Switzerland
[2] ETHZ: Swiss Federal Institute of Technology Zurich, Department of Architecture, Zurich, Switzerland
[3] Alexandria University, Faculty of fine arts, Department of Architecture, Alexandria, Egypt

*(joaopaulo.leitao@eawag.ch, zaghloul@arch.ethz.ch, moosavi@arch.ethz.ch)*

**Abstract:** Physically-based overland flow models are computationally demanding, hindering their use for real-time applications. Therefore, the development of fast (and reasonably accurate) overland flow models is needed if they are to be used to support flood mitigation decision making. In this study, we investigate the potential of Self-Organizing Maps to rapidly generate water depth and flood extent results. To conduct the study, we developed a flood-simulation specific SOM, using cellular automata flood model results and a synthetic DEM and inflow hydrograph. The preliminary results showed that water depth and flood extent results produced by the SOM are reasonably accurate and obtained in a very short period of time. Based on this, it seems that SOMs have the potential to provide critical flood information to support real-time flood mitigation decisions. The findings presented would however require further investigations to obtain general conclusions; these further investigations may include the consideration of real terrain representations, real water supply networks and realistic inflows from pipe bursts.

## 1. INTRODUCTION

Physically-based overland flow models can be used, among other objectives, to support flood risk assessment and management. In urban areas, due to the large number of urban features, such as buildings, roads and also other smaller urban elements, spatial resolution of the terrain representation needs to be high to allow for accurate representation of overland flow characteristics (Leitão et al., 2009). The need for terrain high spatial resolution in urban areas makes these models computationally demanding due to the large number of terrain grid elements, hindering their use for real-time applications, such as urban flood forecasting and real-time urban flood mitigation decision making. To address the issue of long simulation times of physically-based flood models, simplified models such as cellular automata type (e.g. Guidolin et al., 2016) and data-driven models (e.g. Carbajal et al., 2017) have been proposed in recent years.

In this study we explore the potential of Self-Organizing Maps (SOMs) for modelling overland flow originated from point source inflows in real-time. The point source inflows can be originated from, for example, water supply pipe bursts. Pipe bursts are difficult to predict, and due to the considerable impact they can have on urban activities and urban elements, fast flood modelling tools are required to support real-time decisions on how to deploy temporary flood mitigation resources, such as sandbag flood protection walls. In order to evaluate the potential of SOMs for real-time flood modelling in urban areas, we perform an evaluation using a synthetic terrain surface presented by Néelz and Pender (2013) and a fast flood model based on cellular automata (Guidolin et al., 2016). We compare the accuracy of the SOM results evaluating water depth and flood extent at the end of simulation.

---
[*] Corresponding author

## 2. SELF-ORGANIZING MAPS AND FLOOD MODELLING

Self-Organizing Map is a data driven modelling method introduced by Kohonen (1982). From a mathematical point of view, SOM acts as a nonlinear data transformation in which data from high-dimensional space is transformed to low-dimensional space (usually a space of two or three dimensions), while the topology of the original high dimensional space is preserved. In addition to dimensionality reduction and visualization applications, SOM can also be used to predict parameter values or dimensions using data of each other parameter through nonlinear approximation functions (Barreto and Souza, 2006). Designing a proper feature set to represent the data of interest is an important challenging when developing SOMs.

In the literature, applications of SOMs to different fields can be found: wind field modelling (Zaghloul, 2017), air pollution monitoring (Moosavi et al., 2015) and river flood forecasting (Chang et al., 2007). In this work, we implemented a SOM to emulate overland flow in a fast manner when considering detailed terrain representation. The training data to train the SOM is generated from a fast cellular automata flood model proposed by Guidolin et al. (2016); the important SOM features are the position of water source and the 3D geometry around each point in the study area. Given these features plus the final water depths in each location, we train a SOM that learns the relationships between the local geometries, position of water source and the final water depth, and is able to generalize these relationships for new positions of water source without extra hydraulic simulations. An interesting aspect of our SOM features is that since we are focused on water depth at each point, by having only few sample simulations with different positions of water sources, we can generate a relatively large amount of data. However, since in our training data sets most of final water depths are 0 m, we decided to train two different SOMs, one for locations with 0 m water depth and one for locations with positive values (water depth values below 0.03 m were converted to 0 m). After training the SOMs, we combined the weight vectors of the two SOMs with K-Nearest Neighbourhood algorithm to predict the water depths of each point for new inflow locations. The SOM was implemented in Wolfram Mathematica (https://www.wolfram.com/mathematica) and in Grasshopper 3D environment (http://www.grasshopper3d.com).

## 3. PREDICTION OF OVERLAND FLOW

We used the Test 2 presented by Néelz and Pender (2013) as the test bed to explore the potential of SOMs to rapidly simulate flood conditions taking into account detailed terrain representations. This test was originally designed to evaluate the capability of flood models to simulate flood extent and water flood depth. In this particular study, we take advantage of the test detailed terrain representation and consider multiple inflow locations to mimic water supply pipe bursts. Twenty nine (N=29) different inflow locations were generated randomly to represent water supply network pipe mid-points (Figure 1). Twenty two out of the total locations were used to train the SOM (red points in Figure 1), and the remaining locations (seven locations) were used for validation (yellow points in Figure 1). The hydrograph of Test 2 (Néelz and Pender, 2013) was used as the inflow at the inflow locations.

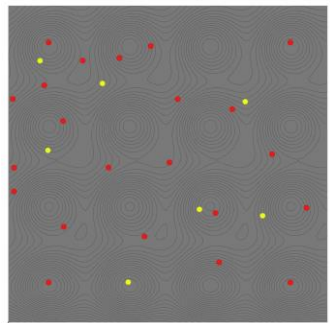

**Figure 1.** Inflow locations representing possible locations of pipe bursts (Red points represent the training inflow locations; Yellow points represent the validation inflow locations; Black lines represent surface contours)

## 4. RESULTS

The flood depth results obtained using the flood model and the SOM for validation locations did not differ significantly. Figure 2 presents graphically the differences between the SOM and flood model results for one of the seven validation configurations. The maximum water depth difference was 0.072 m.

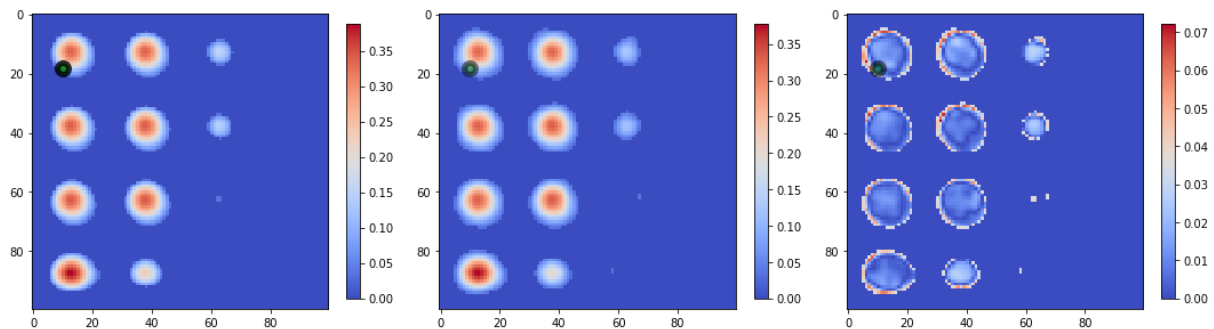

**Figure 2.** Flood depth results and absolute differences between SOM and flood depth model at the end of simulation. Left: flood model; Middle: SOM; Right: absolute differences between SOM and flood model results. The green-black point in the three panels represent the inflow location

For the specific case presented in this study, training the SOM using the 22 training events lasted approximately one hour on a conventional desktop computer (Intel Core i7 CPU, 8 GB RAM), without considering parallel threading. Computation time of flood model and SOM is presented in Table 1 alongside with water depth statistics for the cellular automata flood model and SOM results and their differences. As can be seen in Table 1, computation time was reduced around five times (80%), while the mean water depth differences between the SOM and the flood model results were relatively small: 0.003 m for the mean absolute error which corresponds to a mean relative absolute error of approximately 10%. It is worth noting that computation time reduction shall improve when the number of terrain grid elements and flood volume being simulated increase. Also, if a larger number of training data sets is used, the water depth and flood extent differences will tend to decrease.

**Table 1.** Summary and comparison of flood model and SOM computation time and water depth results. Values in brackets represent relative differences

|  | Computation time | Water depth | | | |
| --- | --- | --- | --- | --- | --- |
|  |  | Minimum | Maximum | Mean | St. Dev. |
| Flood model | (approx.) 11 s | 0 m | 0.389 m | 0.024 m | 0.067 m |
| SOM | (less than) 2 s | 0 m | 0.378 m | 0.025 m | 0.064 m |
| Model results differences | -9 s (-80.0%) | 0.000 m (0.0%) | 0.072 m (159.0%) | 0.003 m (10.7%) | 0.009 m (16.6%) |

## 5. CONCLUSIONS

We have investigated the potential of SOMs to simulate overland flow and flooding caused by point source inflows, such as water supply pipe bursts, using a detailed terrain representation case study. To our knowledge this is the first time that SOMs are used to simulate two-dimensional urban flooding, specifically when caused by point source inflows. In a direct comparison, the results obtained with the SOM were obtained in a small time fraction and were similar to those produced by the fast cellular automata flood model, despite the relatively small number of SOM training events. This suggests that SOMs have the potential to be used for (i) real-time flood forecasting and also for (ii) supporting emergency flood mitigation decision making. Due to their fast computation times, these models can also be used for simulating hazardous substances spills and inform emergency services about the most effective mitigation actions. The findings presented here require further investigations to obtain general conclusions; these further investigations may include the consideration of real (larger and more complex) terrain representations, real water supply networks (topology) and realistic (e.g. time varied) inflows from pipe bursts.